\documentclass[%
 reprint,
superscriptaddress,
 amsmath,amssymb,
 aps,
]{revtex4-2}
\usepackage{graphicx}
\usepackage{dcolumn}
\usepackage{bm}
\usepackage{float}
\usepackage{amsmath}
\usepackage[english]{babel}
\usepackage[autostyle]{csquotes}
\usepackage{multirow} 
\usepackage[flushleft]{threeparttable}
\usepackage{siunitx}
\usepackage{upgreek}
\DeclareMathOperator{\arcsech}{arcsech}

\begin{document}

\title{Variable Frequency Pulse Generation from Breathers in Josephson Transmission Lines}

\author{Gregory Cunningham}
\affiliation{Harvard John A. Paulson School of Engineering and Applied Sciences, Harvard University, Cambridge, MA 02138, USA}
\affiliation{Research Laboratory of Electronics, Massachusetts Institute of Technology, Cambridge, MA 02139, USA}

\author{Yufeng Ye}
\affiliation{Research Laboratory of Electronics, Massachusetts Institute of Technology, Cambridge, MA 02139, USA}

\author{Kaidong Peng}
\affiliation{Research Laboratory of Electronics, Massachusetts Institute of Technology, Cambridge, MA 02139, USA}

\author{Alec Yen}
\affiliation{Research Laboratory of Electronics, Massachusetts Institute of Technology, Cambridge, MA 02139, USA}
\affiliation{Department of Electrical Engineering and Computer Science, Massachusetts Institute of Technology, Cambridge, MA 02139, USA}

\author{Jessica Kedziora}
\affiliation{Research Laboratory of Electronics, Massachusetts Institute of Technology, Cambridge, MA 02139, USA}
\affiliation{Department of Electrical Engineering and Computer Science, Massachusetts Institute of Technology, Cambridge, MA 02139, USA}
\affiliation{MIT Lincoln Laboratory, Lexington, MA 02421, USA}

\author{Kevin P. O'Brien}
\affiliation{Research Laboratory of Electronics, Massachusetts Institute of Technology, Cambridge, MA 02139, USA}
\affiliation{Department of Electrical Engineering and Computer Science, Massachusetts Institute of Technology, Cambridge, MA 02139, USA}

\date{\today}

\begin{abstract}
Single flux quantum technology has the potential to enhance readout and control of superconducting quantum systems due to their low energy consumption, high speed, and cryogenic operating temperatures. Current cryogenic readout and control typically requires microwave pulses of specific frequencies to travel between the room temperature control electronics and the cryogenic setup. Latency in control and readout can be improved by generating pulses within the dilution refrigerator. In this work, we consider a protocol for generating gigahertz frequency microwave tones from trains of DC-centered fluxons and fluxoids in Josephson transmission lines using the dynamics of breather formation, without room temperature synthesis or shunt / bias resistors. Simulations show that pulses with frequencies in the range of 15.2 to 21.5 GHz can be generated with maximal energy efficiency of 97\% and bandwidth from 40 to 365 MHz. This protocol can also be used to generate gigahertz frequency Gaussian pulses. We detail metrics relevant to the control and readout of quantum systems such as input power, output power, and footprint.
\end{abstract}

\maketitle


\section{\label{sec:level1} Introduction}

Scalable superconducting quantum systems provide a promising path towards fault-tolerant quantum information processing. Qubits within today's systems require low-noise microwave signals from control electronics to control and readout quantum states. These signals are typically generated at room temperature and transferred to and from the quantum processor through cables, attenuators, and amplifiers. When optimized for heat load and footprint, industry scale setups, such as IBM's Condor processor, can host upwards of 1100 qubits, though thousands or millions of qubits are required to realize fault tolerance \cite{gambetta2020ibm} \cite{krinner_engineering_2019}.

Single flux quantum (SFQ) technology, operating at or below the 4 K stage of the dilution refrigerator (DR), is an attractive solution to reducing latency in quantum information systems. SFQ control hardware is fast, compared to the coherent evolution of qubits, and power dissipation is proportional to the operation frequency and critical currents of overdamped Josephson junctions (JJs) \cite{savin_high-resolution_2006} \cite{intiso_rapid_2006}. Niobium trilayer Josephson junction fabrication techniques \cite{tolpygo_fabrication_2015} allow for critical current densities down to 1 $\upmu$A/$\upmu$m\textsuperscript{2}, resulting in critical currents in the range of 0.5 - 5 $\upmu$A for integration with superconducting qubits. Heating can be further reduced with coordinated phase evolution with long Josephson junctions (LJJs) \cite{wustmann_reversible_2020}\ or the exclusion of shunt and bias resistors at the JJs, at the cost of ringing behavior from reduced damping \cite{yan_low-noise_2021} \cite{kirichenko_zero_2011}.

Conventional SFQ hardware achieves classical logic and control operations by utilizing propagating flux soliton (fluxon) pulses traveling along Josephson transmission lines (JTL) with current biased and resistively shunted JJs, where the magnetic flux from the pulse is equal exactly to the magnetic flux quantum ($\Phi_{0}$) \cite{Likharev_RSFQ_logic}. Readout via fluxons has been realized with flux qubits \cite{fedorov_fluxon_2014}, and fluxons have been used to coherently control transmon qubits \cite{Leonard_SFQ_Control}\cite{mcdermott_quantumclassical_2018}. Uniquely terminated JTLs without bias or resistively shunted JJs can host breathers, bound state of a fluxon and an antifluxon, that oscillate within the JTL or at the interface of the JTL and termination \cite{mclaughlin_perturbation_1978} \cite{olsen_reflection_1981}. The frequency of the breather oscillation is set by JJ parameters, and in the latter case the oscillations decay based on the degree of impedance mismatch between JTL and termination. 

In this paper we describe the dynamics of breather formation from a single fluxon and engineer a new protocol for generating Gaussian and flat-top Gaussian pulses in the gigahertz range using breather decay from input fluxoid-pair pulse sequences, where fluxoids have a magnetic flux not equal to $\Phi_{0}$. As the simulated JTL system consists of unshunted JJs with no current bias that can be operated at or below 4 K, this microwave pulse generation scheme is an attractive path to reducing the latency of control and readout signals to and from the quantum processor.

\section{\label{sec:level1} Fluxons, The Josephson Transmission Line, and Breathers}

\begin{figure*}[t!]
     \includegraphics[width = 13cm]{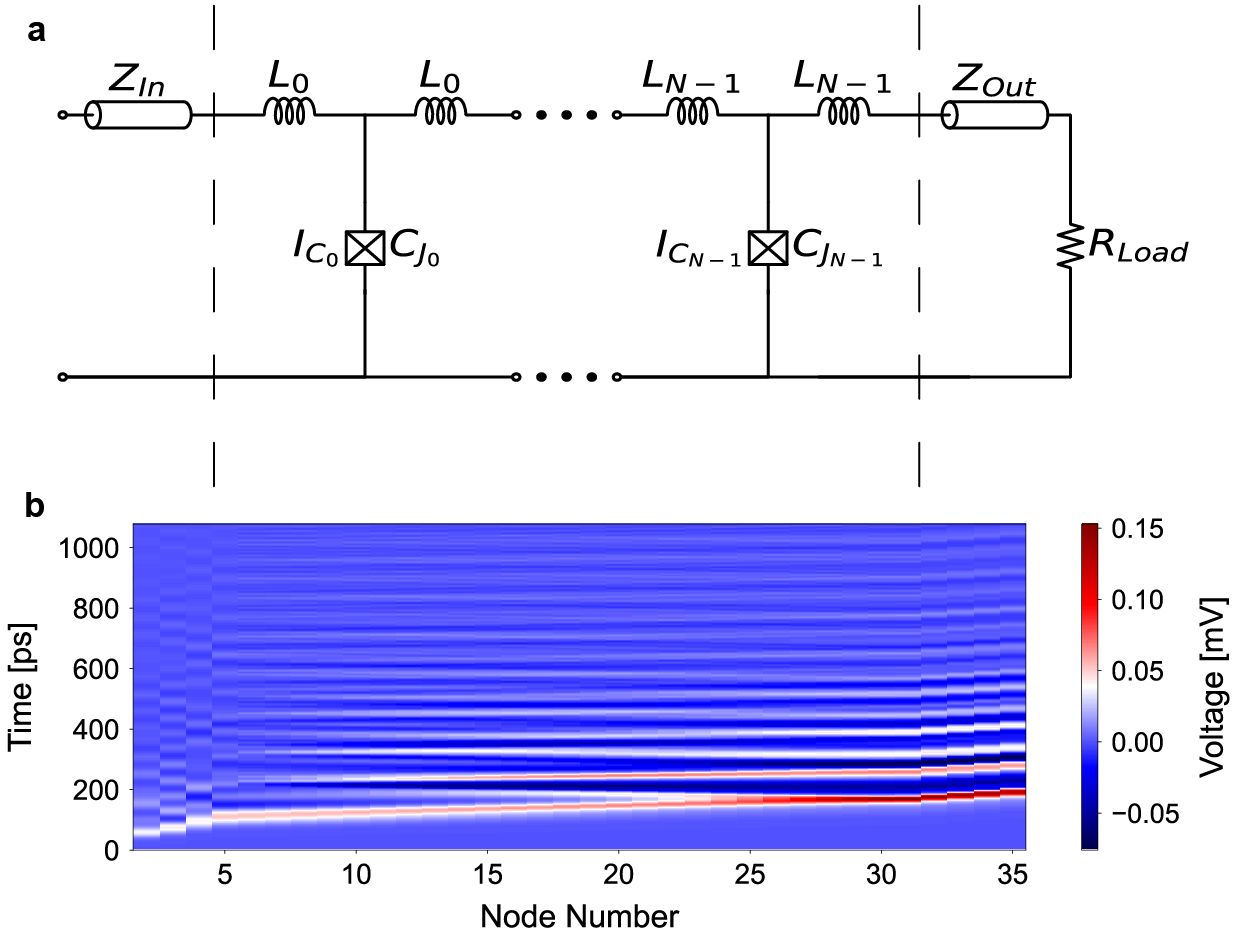}
     \caption{Circuit description of system and simulated breather decay. (a) N unit cell JTL schematic for the nonlinear system described in this work. Critical current ($I_{C}$), junction capacitance ($C_{J}$), and linear inductance ($L$) are the same for each unit cell. JTL impedance ($Z_{JTL}$) is set by $L$ and $C_{J}$. (b) WRspice simulation showing breather formation and decay in a 13 unit cell JTL system with $\lambda_{J}$ = 3.3 unit cells, $\omega_{P}/2\pi$ = 20 GHz, $I_{C}$ = 4 $\upmu$A, $\alpha_{Out} = Z_{JTL} / Z_{Out}$ = 0.2, and $\alpha_{In} = Z_{JTL} / Z_{In}$ = 5.0; time increases vertically (up) and space increases horizontally (right). The fluxon enters the JTL at the first unit cell (Node 5) and propagates with associated plasma radiation towards the final unit cell (Node 30), shown as the red line from t $\approx$ 100 ps to t $\approx$ 180 ps. Due to lack of energy injection from external current bias, a breather (bound state of fluxon and  virtual antifluxon) is formed at the end of the JTL \cite{costabile_exact_1978}. The breather, initially shown as an increase in voltage around 180 ps, oscillates in time whilst temporally and spatially decaying. Damping decreases near the end of the JTL as a result of the output impedance mismatch ($\alpha_{Out}$). After multiple oscillation periods, the breather decays into plasma radiation within the JTL \cite{christiansen_reflection_1980} and all energy from the entire process is dissipated through the input and load terminations.}
     \label{fig:JTL_System_Colorplot}
\end{figure*}

In this section we review the physics of traveling wave solitons in nonlinear media and introduce key parameters and regimes for breather formation. Traveling wave solitons are unique in that the balance of dispersive and nonlinear interactions in their associated media lead to self-reinforcement, allowing the soliton to propagate with a constant shape and velocity as a localized wave \cite{Scott_solitons_1973}\cite{ZS71}. Shape and velocity are maintained upon non-destructive collision with the inclusion of a 2$\pi$ phase shift, whereas destructive collision can result in the formation of two soliton bound state known as a breather that oscillates in space-time \cite{mclaughlin_perturbation_1978} \cite{costabile_exact_1978}. Flux solitons (fluxons) are vortices of magnetic flux equal to exactly $\Phi_{0}$, and are represented in circuits as a hyperbolic secant voltage pulse with time integral area equal to $\Phi_{0}$. 

The standard nonlinear medium that admits fluxons and breathers is the long Josephson junction (LJJ), where the screening length along the long dimension is the Josephson penetration depth ($\lambda_{J} = a \sqrt{L_{J} / L}$, with $a$ as the unit cell length). $\lambda_{J}$ also sets the effective fluxon width \cite{Likharev_RSFQ_logic} \cite{Likharev_JJ_dynamics}. Other defining parameters are the the junction plasma frequency ($\omega_{P} = 1/\sqrt{L_{J} C_{J}}$) \cite{Likharev_JJ_dynamics}, Swihart velocity ($\bar{c} = \lambda_{J} \omega_{P}$) \cite{Likharev_JJ_dynamics} \cite{remoissenet_waves_1996}, and Stewart-McCumber parameter ($\beta_{C} = C_{J}R^2/L_{J}$) \cite{McCumber_OG} \cite{Stewart_OG}. Phase movement across the LJJ is governed by the continuous sine-Gordon equation given by (\ref{LJJ_EoM}). Here subscript \textit{x} denotes spatial derivative and dots above the variable denote temporal derivatives. 

The JTL is the distributed version of the LJJ, comprised of multiple JJs inductively coupled in parallel as shown in Fig.~\ref{fig:JTL_System_Colorplot}a. JTLs become good approximations of LJJs in the limit of high discreteness, where the linear inductance per unit cell is comparable to the Josephson inductance per unit cell ($\lambda_{J} \sim 1$). JTLs are further distinguished from LJJs in that a fluxon traveling in a JTL is accompanied by dissipative plasma radiation, a direct result of the discretization of the sine-Gordon equation (\ref{JTL_EoM}) governing phase movement in either system \cite{mclaughlin_perturbation_1978} \cite{kivshar_dynamics_1989} \cite{peyrard_kink_1984} \cite{kurin_radiation_1997}. Except for one special logic type which uses coupled LJJs for ballistic computing, JTLs used for digital logic are typically discrete and use a current bias of $\sim 70\%$ of $I_{C}$ at each junction to counteract this energy loss process \cite{Likharev_RSFQ_logic} \cite{osborn_asynchronous_2022}.

\begin{subequations}\label{LJJ_JTL_EoMs}
\begin{align}
    \sin\phi - \lambda_{J}^2 \phi_{xx} + \omega_{P}^{-2} \ddot\phi = 0 \label{LJJ_EoM} \\
    \sin\phi_{n} - \lambda_{J}^2(\phi_{n+1} + \phi_{n-1} - 2\phi_{n}) + \omega_{P}^{-2} \ddot\phi_{n} = 0 \label{JTL_EoM} 
\end{align}
\end{subequations}

We now describe propagation of fluxons across JTL unit cells in a low discreteness (quasi-continuous) regime. We will show in later sections that this sets necessary bounds on input parameters. Propagation occurs via the phase slip process \cite{Likharev_RSFQ_logic}, where a fluxon pulse incident on a unit cell will lower the Josephson potential barrier between equilibria and cause the potential landscape to tilt, resulting in a \enquote{washboard} shape that the phase will roll down \cite{Likharev_JJ_dynamics} \cite{darula_dynamic_1990}. In the process of rolling, a voltage equal to that of the fluxon pulse builds up across the JJ in the given unit cell and travels to the next via the linear inductance ($L$). The potential landscape in the first unit cell goes back to the standard cosine configuration, and the phase oscillates about the new equilibrium $2\pi$ radians away from the original equilibrium with damping determined by the ratio of relaxation timescales, $\beta_{C}$. In our overdamped system ($\beta_{C} << 1$), the relaxation between the nonlinear inductor and normal state resistance ($\tau_{LR} = L_{J} / R$) sets the timescale for this phase slip process. Fluxon dynamics can be inferred based on the scaled velocity ($\widetilde{v} = v / \bar{c}$) and the fluxon effective width ($W = \lambda_{J}\sqrt{1- \widetilde{v}^2}$) \cite{wustmann_reversible_2020}, where the former controls how much kinetic energy the fluxon will have when encountering the Josephson potential barrier \cite{peyrard_kink_1984} \cite{currie_numerical_1977}  and the latter controls how the built-up voltage spreads among unit cells \cite{currie_numerical_1977} \cite{polonsky_transmission_nodate}.

\begin{figure}[t!]
     \includegraphics[scale = 0.25]{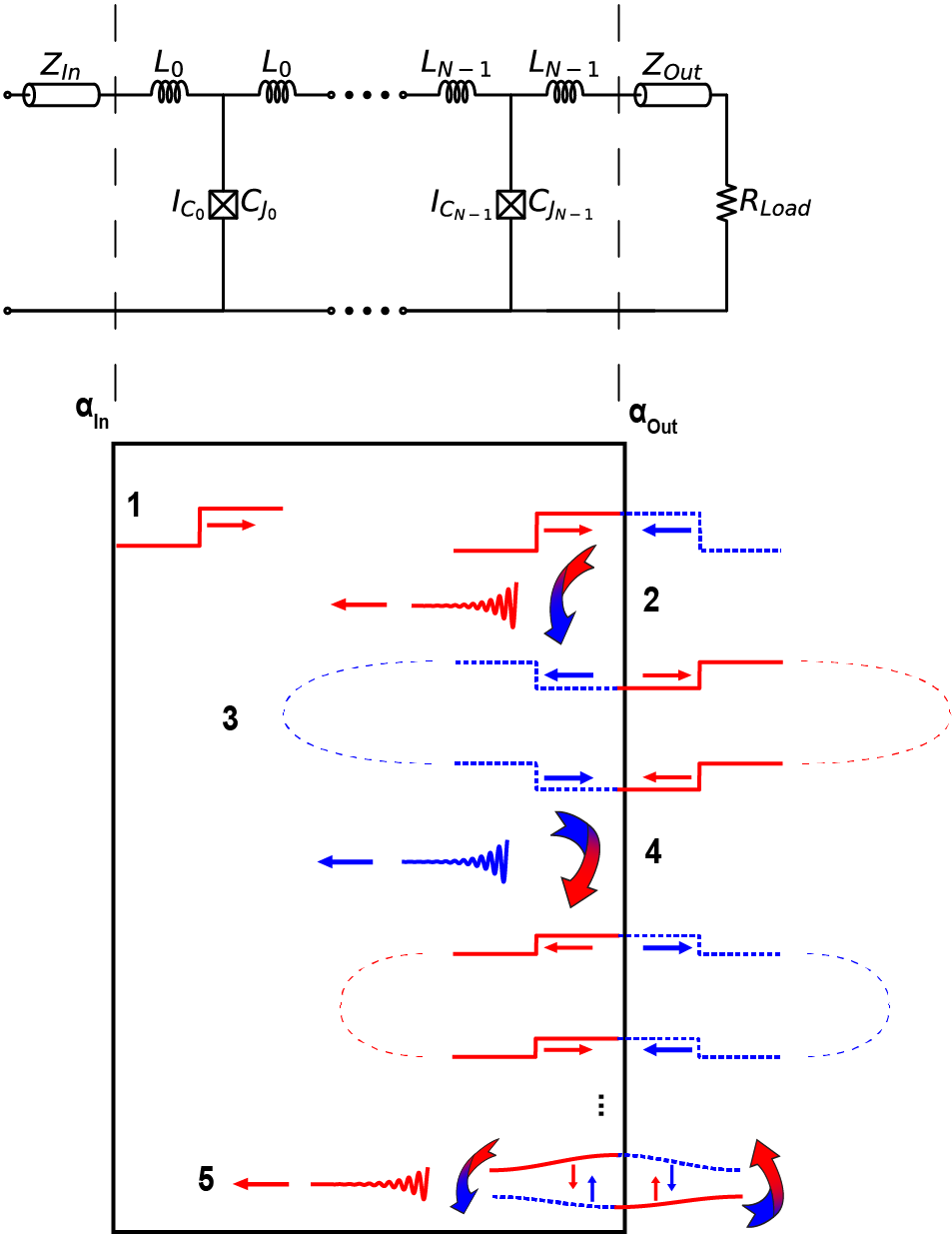}
     \caption{Toy model for breather formation and decay from fluxon within JTL. 1.) A $2\pi$ fluxon (red) enters the JTL and travels to the end of the line. 2.) Fluxon approaches the output transmission line ($\alpha_{Out} \sim > \alpha_{0}$), leading to a collision between fluxon and virtual antifluxon traveling in opposite direction (dashed blue). THe collision forms a breather (bound fluxon pair state) with some plasma radiation that travels towards the input ($\alpha_{0} < \alpha_{In} << \alpha_{\infty}$). 3.) Having lost energy during formation (shown as smaller amplitude fluxoid / antifluxoid), the bound pair can not completely separate in space-time and they travel back towards the boundary. 4.) The process of decaying oscillation about the boundary continues for the breather, with each polarity change accompanied by plasma radiation. 5.) After multiple oscillation periods, the breather amplitude is too small support the fluxoid bound state and the breather decays into plasma radiation. \cite{mclaughlin_perturbation_1978} \cite{costabile_exact_1978}}
     \label{fig:breather_toy_model}
\end{figure}

Having covered propagation within the JTL, boundary conditions at either port determine the formation and trapping of breathers that will generate the desired microwave frequency pulses. Current conservation at a boundary simplifies to $\phi_{X} - \alpha\phi_{T} = 0$ after introduction of dimensionless variables ($X = x/\lambda_{J}, T = t\omega_{P}$). The impedance ratio between the JTL and termination ($\alpha = \sqrt{\frac{L}{C_{J}}}/Z$) determines the reflection dynamics. For fixed $\widetilde{v}$, reflection at the boundary can be considered as a collision between a forward propagating real fluxon and a backward propagating virtual antifluxon (fluxon), a process which will increase (decrease) the phase by 2$\pi$ \cite{mclaughlin_perturbation_1978} \cite{costabile_exact_1978}.

\begin{figure*}[t!]
     \centerline{\includegraphics[scale = 0.34]{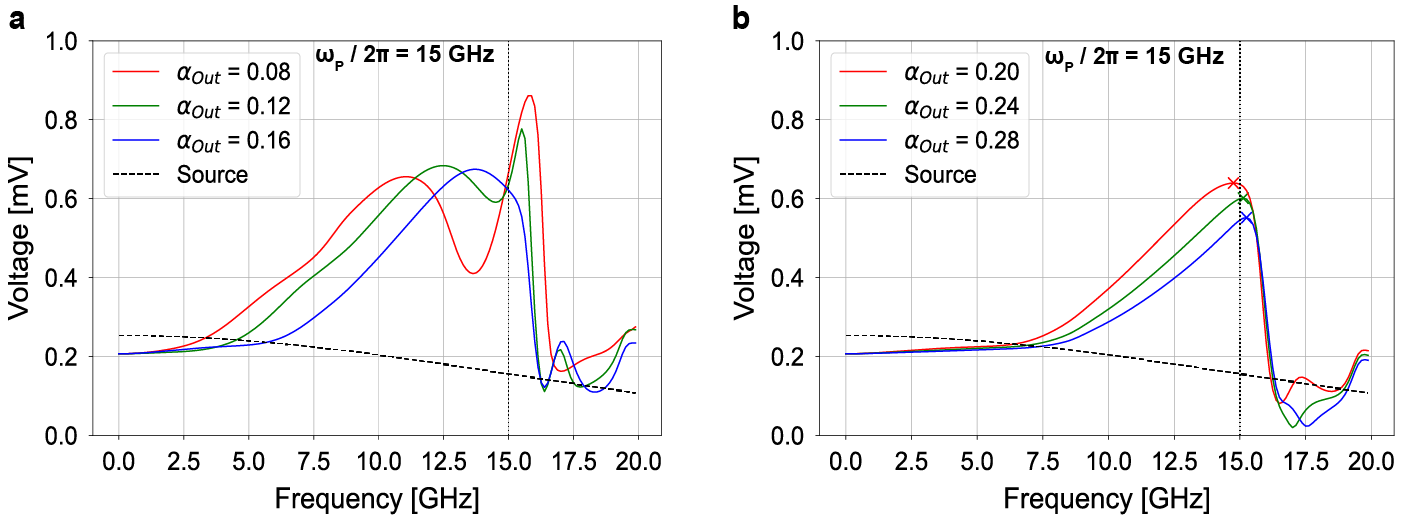}}
     \caption{Breather variation with output impedance mismatch ($\alpha_{Out}$) for fixed fluxon input (dashed line). (a) Frequency domain picture of breather decay for fixed $\omega_{P}$ (dotted line) in JTL system for $\alpha_{Out} \sim \alpha_{0} = 0.075$. In this range, the energy loss during reflection is small enough for the real (antifluxon) and virtual (fluxon) components to separate at the JTL - output interface with plasma radiation near $\omega_{P}$, leading to stronger components at lower frequencies. (b) Frequency domain picture of breather decay for fixed $\omega_{P}$ (dotted line) in JTL system for $\alpha_{Out} > \alpha_{0} = 0.075$. In this range, the real (antifluxon) and virtual (fluxon) components do not have enough energy to separate after initial reflection, leading to breather formation. Resonance shifts upward with increasing $\alpha_{Out}$ in agreement with theory \cite{costabile_exact_1978}. Resonance for each $\alpha_{Out}$ are marked with x. JTL parameters for (a) and (b): length = 13 unit cells, $\lambda_{J}$ = 3.3 unit cells, $I_{C}$ = 4 $\upmu$A, $\omega_{P} / 2\pi$ = 15 GHz.} 
     \label{fig:Impedance_ratio_variation}
\end{figure*}

The approximate solution at the boundary will be a state combining the real and virtual waves, where the simplest case has both waves separate from each other as t approaches $\pm \infty$ \cite{mclaughlin_perturbation_1978}. This assumes that the energy of the combined state is greater than or equal to the rest energy of the real and virtual components, which would be twice the rest energy of a fluxon ($2E_{0}$, $E_{0} = 8E_{J}\lambda_{J}$) \cite{olsen_reflection_1981}. The thresholds where this assumption holds are set by the boundary impedance mismatch as a function of the incident fluxon kinetic energy (velocity), and are given by $\alpha_{0}$ (\ref{antifluxon_reflection}) and $\alpha_{\infty}$ (\ref{fluxon_reflection}). For $\alpha < \alpha_{0}$ ($\alpha > \alpha_{\infty}$), the incident fluxon reflects as an antifluxon (fluxon) with noticeable (minimal) plasma radiation near $\omega_{P}$. For $\widetilde{v} = 0.75$, $\alpha_{0} = 0.075$ and $\alpha_{\infty} = 4.54$. However, due to losses at the boundary and lack of energy injection in the current system, the combined state will have energy less than $2E_{0}$, leading to the absorption regime ($\alpha_{0} \leq \alpha \leq \alpha_{\infty}$) where the fluxon is absorbed by the termination with some reflection or a breather is formed \cite{wustmann_reversible_2020}. Breathers oscillate at frequencies near $\omega_{P}$ and have a temporal decay constant inversely proportional to $\alpha$ \cite{mclaughlin_perturbation_1978}.

\begin{subequations}\label{Impedance_Ratio_Conditions}
\begin{align}
    \alpha_{0}(\widetilde{v}) = \left| \frac{\sqrt{1 - \widetilde{v}^{2}} - 1}{2(\arctan(\sqrt{1-\widetilde{v}^{2}}/\widetilde{v}) / \sqrt{1 - \widetilde{v}^{2}} + \widetilde{v})} \right|\label{antifluxon_reflection} \\
    \alpha_{\infty}(\widetilde{v}) =  \left| \frac{4\widetilde{v}}{\sqrt{1 - \widetilde{v}^{2}} - 1} \right|\label{fluxon_reflection}
\end{align}
\end{subequations}

\begin{figure*}[t!]
     \includegraphics[width = 17cm]{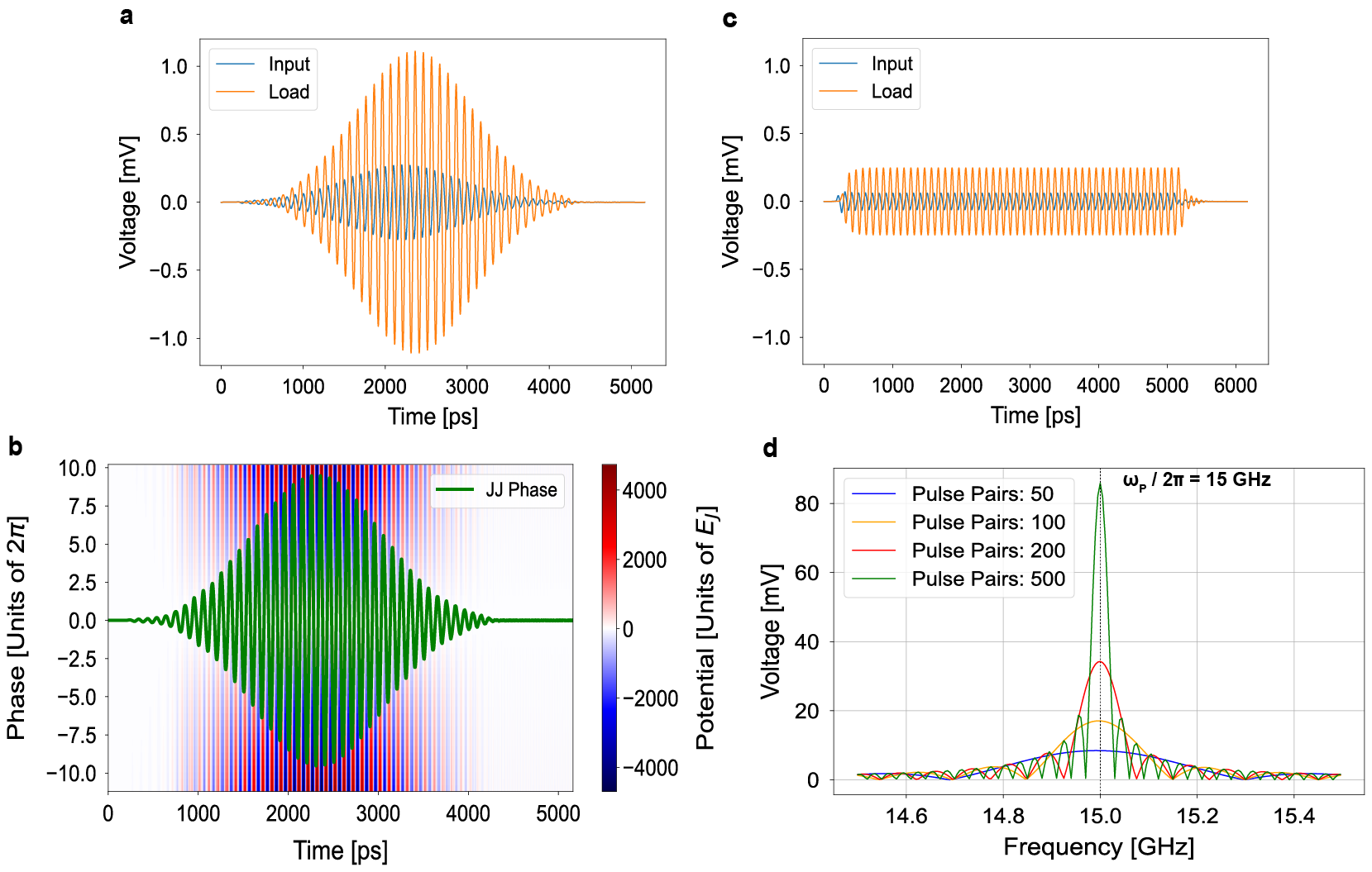}
     \caption{Gaussian and flat-top Gaussian pulse generation. (a) Gaussian load (orange) transient resulting from input (blue) of 41 pulse pairs of fluxoids and antifluxoids (4.045 ns sequence duration) whose amplitudes follow a Gaussian distribution. (b) Temporal propagation of JTL unit cell potential energy (red/blue/white) and junction phase (green) after application of 41 pulse pairs for Gaussian generation. Using a spacing of a breather period ($\omega_{P}^{-1}$) between pulse pairs creates a periodic potential for the junction phase, with maximal phase accrued during oscillation determined by the amount of flux contained within the fluxoid or antifluxoid. JTL parameters for (a) and (b): length = 4 unit cells, $\lambda_{J}$ = 2.50 unit cells, $I_{C}$ = 3.0 $\upmu$A, $\alpha_{Out}$ = 0.25, $\omega_{P} / 2\pi$ = 10 GHz. (c) Flat-top Gaussian load (orange) transient resulting from input (blue) of 50 pulse pairs of fluxons and antifluxons (5.138 ns sequence duration). All pulses have uniform amplitude aside from the initial and final pulses such that the final equilibrium phase is 0 radians. JTL parameters: length = 5 unit cells, $\lambda_{J}$ = 3.17 unit cells, $I_{C}$ = 4.0 $\upmu$A, $\alpha_{Out}$ = 0.25, $\omega_{P} / 2\pi$ = 10 GHz. (d) Bandwidth narrowing of flat-top Gaussian pulse in frequency domain for various numbers of pulse pairs with a fixed $\omega_{P}$ (dotted line) JTL. Increasing the number of pulse pairs enhances and maintains the periodic potential for longer time scales, achieving bandwidths of 40 MHz for a 500 pulse pair sequence. Parameters: pulse width = 30.71 ps, pulse spacing = 100 ps, JTL length = 5 unit cells, $\lambda_{J}$ = 3.17 unit cells, $I_{C}$ = 3.0 $\upmu$A, $\alpha_{Out}$ = 0.25, $\omega_{P} / 2\pi$ = 15 GHz.}\label{fig:transients_Ujtl_BWcomp}
\end{figure*}

A WRspice simulation of breather formation and decay from a propagating fluxon in the unbiased-terminated JTL system is shown in Fig.~\ref{fig:JTL_System_Colorplot}b, with $\alpha_{In} = 5$ and $\alpha_{Out} = 0.2$  for $R_{Load}$ = 15.7 $\Upomega$. A toy model of the breather formation process is described in Fig.~\ref{fig:breather_toy_model}. An oscillating breather is formed, with associated plasma radiation, from the fluxon collision at the load and decays spatially and temporally within the JTL until all energy is dissipated through the terminations. Note that the input impedance ratio is optimized for fluxon transmission into the JTL ($\alpha_{0} < \alpha_{In} = Z_{JTL} / Z_{In} << \alpha_{\infty}$) while the load impedance ratio is set for breather formation and trapping within the JTL ($\alpha_{Out} = Z_{JTL} / Z_{Out} \sim > \alpha_{0}$). It should be noted that breather reflection at a boundary follows similar trends as fluxon reflection, where $\alpha$ approaching 0 reflects the breather with plasma radiation near $\omega_{P}$ and $\alpha$ approaching $\infty$ reflects the breather with minimal plasma radiation \cite{olsen_reflection_1981}.

\section{\label{sec:level1} Single Fluxon: Breather Decay Simulations}

This section details the required input and JTL operating ranges for breather formation as confirmation of the above theory. To allow for breather decay, the fluxon must first cleanly propagate through the unbiased JTL and reach the output termination with $\alpha_{Out} \gtrsim \alpha_{0}$. We set the product of the JTL inductance and junction critical current to be much less than a flux quantum to avoid Cherenkov radiation \cite{mclaughlin_perturbation_1978} and flux trapping in a given unit cell \cite{polonsky_transmission_nodate}, a condition equivalent to $\lambda_{J} >>$ 1 unit cell. We also set the input transmission line impedance to be much smaller than the JTL impedance to minimize fluxon reflection at the input port (leftmost dashed line in Fig.~\ref{fig:JTL_System_Colorplot}a) during forward propagation, leading to an input impedance ratio ($\alpha_{In}$) of 5. Finally, we set the fluxon pulse width based on half of the maximum amplitude ($2W\arcsech(0.5)/v_{0}$) for efficient phase slip propagation between unit cells and fix $\widetilde{v_{0}} = 0.75$ \cite{mclaughlin_perturbation_1978} \cite{remoissenet_waves_1996}.

The proposed system utilizes input and output transmission lines whose electrical lengths are small compared to the product of the fluxon pulse width and $\bar{c}$, with the input and output lengths equal to $0.4 \bar{c}\Phi_{0}/I_{C}R_{N}$ unit cells \cite{Likharev_RSFQ_logic}. The JTL length for single fluxon simulations is set at $4\lambda_{J}$ unit cells.

Within the absorption region, breather formation occurs as the impedance ratio between the JTL and output transmission line decreases from unity ($\alpha_{Out} < 1$). WRspice simulations of different breather regimes after formation are shown in Fig.~\ref{fig:Impedance_ratio_variation}. For $\alpha_{Out} \sim \alpha_{0} = 0.075$ we operate on the edge of the absorption regime, where reflection at the interface yields an antifluxon. Noticeable plasma radiation near $\omega_{P}$ occurs as we increase $\alpha_{Out}$, as shown by the blue line in Fig.~\ref{fig:Impedance_ratio_variation}a \cite{costabile_exact_1978}. As we head further into the absorption region by increasing $\alpha_{Out}$, breather formation occurs with the distinguishing resonance shift and magnitude decrease as shown in Fig.~\ref{fig:Impedance_ratio_variation}b \cite{costabile_exact_1978}. The regime we will focus on for subsequent analysis is that for which $\alpha_{Out}$ is between 0.35 and 0.15.

\section{\label{sec:level1} Multiple Fluxons and Fluxoids: Flat-Top Gaussian and Gaussian Pulse Generation}

In this section, we build upon the results of breather formation from single fluxons and detail how quantities such as bandwidth and generated frequency can be enhanced with trains of fluxons and fluxoids, which are defined as voltage pulses with time integral area not equal to $\Phi_{0}$ but whose phase winding quantization remains valid \cite{fluxoids_ustinov}. Sending multiple fluxons or fluxoids into the JTL can create pulse shapes formed by decaying breather oscillations about equilibrium. Patterns arise by timing the pulses to alter the JTL potential landscape at half periods of the breather ($T_{B} \sim \omega_{P}^{-1}$), causing the JJ phase to oscillate. The oscillation amplitudes are determined by the amount of flux introduced by each pulse, where a flux of +/- $\Phi_{0}$ will change the phase by +/- $2\pi$. Each pulse is followed by a pulse of opposite polarity with sufficient flux to keep the same phase magnitude but change the sign. The pulses have fixed pulse widths equal to the largest relaxation timescale in an overdamped junction ($\tau_{LR}$) and the input / output transmission lines have length $0.4 \bar{c}\Phi_{0}/I_{C}R_{N}$ unit cells. Additionally, the JTL length is set to minimize the input impedance seen after the input port. For our given range of $\alpha_{Out}$, a traveling wave with frequency $\omega_{P}$ sees minimal input impedance when $\sim |1/i\tan(N_{JTL}/\lambda_{J})|$ is minimized. The first minimum at $N_{JTL}/\lambda_{J} \sim 1.66$ sets the JTL length.

Gaussian pulse generation is a useful application of this system, as these specific pulses are typically used in superconducting qubit readout \cite{krantz_quantum_2019}. Input and load transients for a 41 pulse pair sequence are shown in Fig.~\ref{fig:transients_Ujtl_BWcomp}a. This is achieved by varying the amount of flux introduced by each pulse to follow a Gaussian envelope pattern. The potential engineered by this pulse sequence causes phase oscillations that follow the same pattern as shown in Fig.~\ref{fig:transients_Ujtl_BWcomp}b. Flat-top Gaussian pulse generation from multiple fluxon pulses is another application of this system tailored toward superconducting qubit readout and control \cite{krantz_quantum_2019}. Input and load transients for a 50 pulse pair flat-top Gaussian sequence are shown in Fig.~\ref{fig:transients_Ujtl_BWcomp}c.

The rationale behind using multiple pulses as opposed to a single pulse is that the bandwidth is significantly smaller, on the order of MHz as opposed to GHz, which is necessary for realistic quantum control and readout. By timing incoming pulses to alter the JTL potential while the phase is near the first extrema during breather oscillation and decay ($\sim$ +/- $2\pi$ for a fluxon / antifluxon), the potential barrier that the phase approaches is decreased by the incoming fluxon (increased by the incoming antifluxon), leading to a strongly parabolic potential landscape. The resulting phase and voltage oscillations are then fixed tighter to the fundamental frequency $\omega_{P}$. This effect is shown in Fig.~\ref{fig:transients_Ujtl_BWcomp}d for flat-top Gaussian generation, where increasing the number of pulse pairs from 50 to 500 decreases the FWHM from 365 MHz to 40 MHz, where the latter is comparable to the FWHM of pulses typically used for transmon qubit control and readout \cite{Leonard_SFQ_Control} \cite{Transmon_OG_Koch}.

\begin{table*}
    \centering
    \caption{Simulated System Performance}
    \begin{tabular}{l c c c c c c c c}
        \hline
        \hline
        Application & I\textsubscript{C} & L & C\textsubscript{J} & N\textsubscript{pulses} & $\upomega_{0} / 2\pi$  &  FWHM & Average Power [Input] & FWHM Power [Load]\\
        
        Flat-Top Gaussian & \SI{3}{\micro A} & \SI{10.903}{\pico H} & \SI{800}{\femto F} & \num{50} & \SI{16.991}{\giga Hz} & \SI{418}{\mega Hz} & \SI{1.860}{\nano W} & \SI{-77.213}{dBm}\\ 
        Flat-Top Gaussian & \SI{4}{\micro A} & \SI{8.177}{\pico H} & \SI{800}{\femto F} & \num{50} & \SI{19.609}{\giga Hz} & \SI{482}{\mega Hz} & \SI{3.893}{\nano W} & \SI{-74.472}{dBm}\\ 
        Flat-Top Gaussian & \SI{5}{\micro A} & \SI{6.542}{\pico H} & \SI{800}{\femto F} & \num{50} & \SI{21.918}{\giga Hz} & \SI{536}{\mega Hz} & \SI{6.475}{\nano W} & \SI{-73.056}{dBm}\\ 
        Flat-Top Gaussian & \SI{6}{\micro A} & \SI{5.453}{\pico H} & \SI{800}{\femto F} & \num{50} & \SI{24.018}{\giga Hz} & \SI{582}{\mega Hz} & \SI{10.135}{\nano W} & \SI{-71.448}{dBm}\\ 
        
        Gaussian & \SI{3}{\micro A} & \SI{17.542}{\pico H} & \SI{1000}{\femto F} & \num{41} & \SI{15.191}{\giga Hz} & \SI{836}{\mega Hz} & \SI{36.207}{\nano W} & \SI{-65.446}{dBm}\\
        Gaussian & \SI{4}{\micro A} & \SI{13.159}{\pico H} & \SI{1000}{\femto F} & \num{41} & \SI{17.536}{\giga Hz} & \SI{964}{\mega Hz} & \SI{56.605}{\nano W} & \SI{-63.957}{dBm}\\
        Gaussian & \SI{5}{\micro A} & \SI{10.527}{\pico H} & \SI{1000}{\femto F} & \num{41} & \SI{19.609}{\giga Hz} & \SI{1073}{\mega Hz} & \SI{72.096}{\nano W} & \SI{-63.308}{dBm}\\
        Gaussian & \SI{6}{\micro A} & \SI{8.773}{\pico H} & \SI{1000}{\femto F} & \num{41} & \SI{21.482}{\giga Hz} & \SI{1164}{\mega Hz} & \SI{97.012}{\nano W} & \SI{-62.402}{dBm}\\
        \hline
        \hline
    \end{tabular}
    \label{tab:Simulated_system_perfomance}
\end{table*}

\section{\label{sec:level1} Multiple Fluxons and Fluxoids: Energy Efficiency}

\begin{figure}[t!]
     \includegraphics[width = 8.6cm]{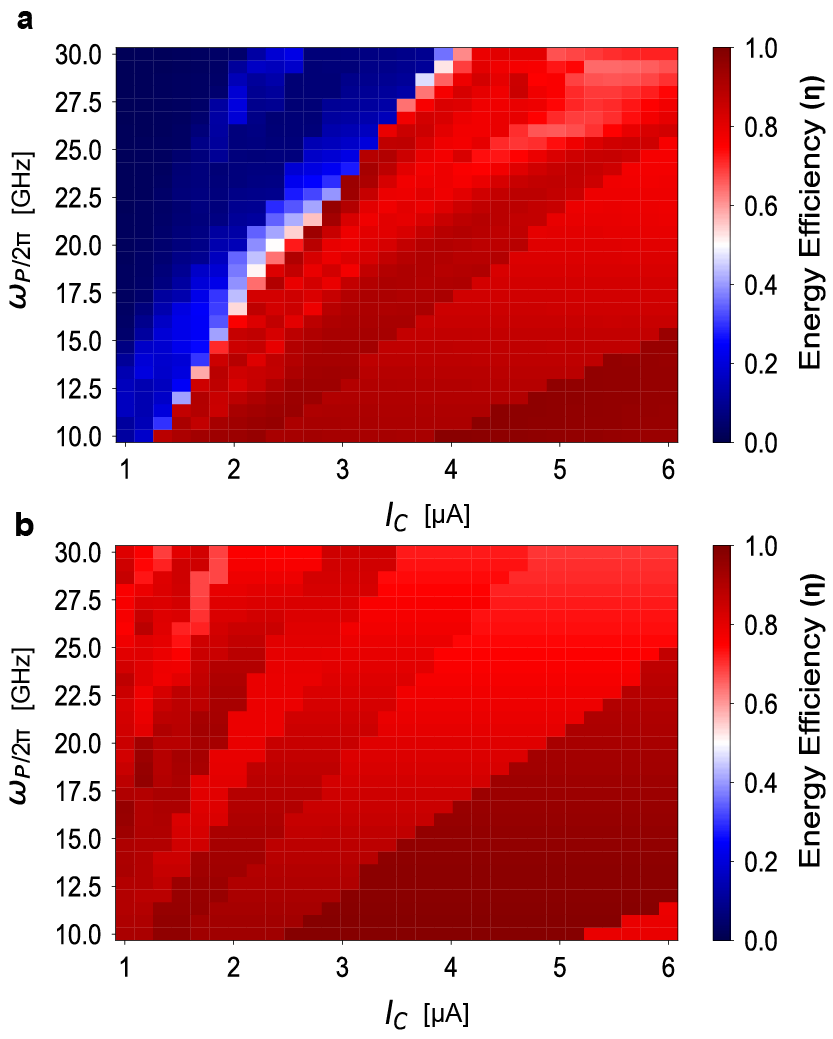}
     \caption{Flat-top Gaussian and Gaussian pulse generation energy efficiency. (a) Effects of $I_{C}$ and $\omega_{P}$ on energy efficiency for flat-top Gaussian generation (50 pulse pairs), $\eta_{Max}$ = 0.970. Phase motion in the first unit cell of the JTL increases with increasing $I_{C}$, leading to greater efficiency as a majority of the pulse sequence propagates through the JTL and contributes to power flow at the load. JTL parameters: length = 5 unit cells, $\lambda_{J}$ = 3.17 unit cells, $\alpha_{Out}$ = 0.25. (b) Effects of $I_{C}$ and $\omega_{P}$ on energy efficiency for Gaussian generation (41 pulse pairs), $\eta_{Max}$ = 0.987. Currents applied to the first unit cell of the JTL from the pulse sequence are much larger than $I_{C}$, increasing phase motion to yield greater efficiency even for small $I_{C}$. JTL parameters: length = 4 unit cells, $\lambda_{J}$ = 2.50 unit cells, $\alpha_{Out}$ = 0.25.}
     \label{fig:Energy_Efficiency}
\end{figure}

Energy efficiency is an important metric for superconducting quantum systems as loss can reduce the amount of power that transfers from one component to the next, hindering operation due to sensitivities in dissipation (destruction of quantum states) or desired input (gain regimes for low temperature amplifiers). We define load energy efficiency ($\eta$) as the ratio of the time integral of power flow out of the system (right of $Z_{Out}$) to the time integral of power flow into system (left of $Z_{In}$). Power flow into and out of the system corresponds to the forward power waves at the input and output ports, respectively, and is dependent on $I_{C}$ and $C_{J}$. 

Using the analogy of a pendulum, the critical current sets the potential energy ($mgl$) and the junction capacitance sets the moment of inertia ($ml^{2}$) \cite{Likharev_RSFQ_logic} \cite{Gronbech_Jensen_fluxon_pendulum}. For small $I_{C}$, there is not enough weight for the pendulum phase to move far from equilibrium when torque is applied. In other words, the current applied to a junction from the pulse will effectively be equal to the current across the junction capacitor ($I_{a} = C_{J}\ddot{\phi}$). With $\ddot{\phi}$ small due to the phase staying near equilibrium, the majority of the forward propagating pulse is canceled by a virtual backward propagating pulse that contributes to the backward power flow through the input transmission line. For $I_{C}$ large enough to avoid this behavior, increasing $C_{J}$ will result in the majority of the forward propagating pulse making it to the next unit cell and eventually contributing to the forward power flow at the output port. Energy efficiency for flat-top Gaussian generation (50 pulse pairs) follows this trend as shown in Fig.~\ref{fig:Energy_Efficiency}a, where small $\omega_{P}$ improves efficiency for a given critical current and large critical currents yield better efficiency over the given range of $\omega_{P}$. 

The situation changes when the current applied to the junction during a pulse is large compared to the critical current. In this case, the junction phase does not stay near equilibrium for small $I_{C}$ and the majority of the forward propagating pulse reaches the end of the line during the pulse sequence. Energy efficiency for Gaussian generation (41 pulse pairs) showcases this behavior in Fig.~\ref{fig:Energy_Efficiency}b, where efficiency is high even for small $I_{C}$. Maximal energy efficiency for flat-top Gaussian and Gaussian generation are 0.970 and 0.987, respectively.

\section{\label{sec:level1} Outlook and Conclusions}

We conclude this work with a few remarks on the JTL system, the input pulse sequence, and applications to superconducting dispersive readout. For the parameter ranges considered in this system, the minimal and maximal JTL lengths for pulse generation will be 4 ($\lambda_{J} = 2.50$) and 5 ($\lambda_{J} = 3.17$) unit cells, respectively. For critical currents within the range of 0.5 - 5 $\mu$A in a 5 unit cell JTL, this corresponds to total junction footprints of 0.5 - 5 $\upmu$m\textsuperscript{2} using standard $J_{C}$ = 1 $\upmu$A/$\upmu$m\textsuperscript{2} processes. The majority of the footprint will be dominated by shunt capacitance, as low GHz tones require total capacitance (shunt + junction) on the order of 1-3 pF for $I_{C}$ above 3 $\upmu$A. As an additional perk, the impedance mismatch between the JTL and output impedance required to form breathers allows for voltage gain in the output pulse compared to pulse generation from the same input pulse train through a passive transmission line with matched load. 

The SFQ pulses in this work would be generated via a DC-SFQ converter as described in \cite{Likharev_RSFQ_logic}. Given the pulse spacing between SFQ pulses in \cite{Leonard_SFQ_Control}
($\sim \omega^{-1}$, with $\omega \leq 2\pi \times$ 5 GHz), we believe generation of the initial pulse sequence is feasible for the frequencies considered in this work (see Table~\ref{tab:Simulated_system_perfomance}). Alternative schemes for SFQ pulse generation such as the Josephson pulse generator \cite{JPG_3K} may not offer the flexibility of generating pulses of opposite polarity one after the other. Increasing the number of pulse pairs for flat top Gaussian generation has the additional benefit of increased energy efficiency. However, energy efficiency saturates near 0.97 for large quantity of pulse pairs, likely due to radiation that arises during breather formation. Additionally, increasing the number of pulse pairs increases the average power at the input port for Gaussian and flat-top Gaussian generation. In our simulations the maximum average power input to the system for flat-top Gaussian generation is 10.135 nW (see Table~\ref{tab:Simulated_system_perfomance}) which is above the power needed for few photon qubit readout ($\sim 1\times10^{-18}$ W) \cite{Gambetta_qubit_readout_measurement} but well below the limit set by the cooling power of the mK stage of the DR ($\sim 19 \upmu$W) \cite{krinner_engineering_2019}.

A potential future direction for this work is reverse operation to generate fluxoids from microwave pulses. With adjustments to operation ranges of $\alpha_{Out}$, $\lambda_{J}$, and the input pulse frequency, the reverse system operating at mK would connect to conventional SFQ logic circuits at 4 K and further reduce the latency in post-processing quantum systems, presenting an alternative path for SFQ-based measurement of qubits compared with measurement systems based on fluxon velocity shifts \cite{fedorov_fluxon_2014}. We imagine a full system consisting of the forward (fluxoids to Gaussian) and reverse (Gaussian to fluxoids) devices operating at 4 K. Readout tones generated by the forward device will be fed to the quantum processor with superconducting qubits and readout resonators at the mK stage. Using a reflection style readout configuration with the readout tone slightly detuned from the resonator frequency, the resultant Gaussian will undergo a phase shift based on the qubit state \cite{krantz_quantum_2019}. A cryogenic IQ mixer \cite{Josephson_IQmixer} can then be used to separate the in-phase and quadrature components, each of which will have a different amplitude envelope. Feeding both into two reverse systems at 4 K will generate two different total fluxes (from the fluxoids and antifluxoids) that can be compared to determine the qubit state, depending on the detuning between the readout pulse and resonator frequencies.

In summary, we have shown the decaying oscillation of breathers from sequences of fluxon and antifluxon pulses in an unbiased JTL system with unshunted JTLs and load impedance mismatch can produce flat-top Gaussian pulses with center frequencies of 15.2 - 21.5 GHz (Table~\ref{tab:Simulated_system_perfomance}) and simulated maximum energy efficiency of 0.97. Increasing the number of pulse pairs can dramatically reduce the bandwidth to as low as 40 MHz, comparable with pulse bandwidths used for qubit readout and control, with output powers ranging from -74.5 to -69.3 dBm. Sequences of fluxoids and antifluxoids can be configured to generate Gaussian pulses with center frequencies of 14.6 - 21.1 GHz and simulated maximum energy efficiency of 0.98. With the ability to operate at both 4 K and mK temperatures due to small critical currents (0.6 - 6 $\upmu$A) and generate tones within and outside of the frequency range of today's qubits, this system may be a viable option for higher frequency / temperature quantum applications.

\section{\label{sec:level1} Acknowledgments}

Gregory Cunningham would like to thank the members of the QCE research group for insightful discussions and assistance with code development. Gregory Cunningham would like to thank Kevin D. Osborn for insightful discussions.

\def\bibsection{\section*{\refname}}

\bibliography{refs}

\begin{thebibliography}{36}%
\makeatletter
\providecommand \@ifxundefined [1]{%
 \@ifx{#1\undefined}
}%
\providecommand \@ifnum [1]{%
 \ifnum #1\expandafter \@firstoftwo
 \else \expandafter \@secondoftwo
 \fi
}%
\providecommand \@ifx [1]{%
 \ifx #1\expandafter \@firstoftwo
 \else \expandafter \@secondoftwo
 \fi
}%
\providecommand \natexlab [1]{#1}%
\providecommand \enquote  [1]{``#1''}%
\providecommand \bibnamefont  [1]{#1}%
\providecommand \bibfnamefont [1]{#1}%
\providecommand \citenamefont [1]{#1}%
\providecommand \href@noop [0]{\@secondoftwo}%
\providecommand \href [0]{\begingroup \@sanitize@url \@href}%
\providecommand \@href[1]{\@@startlink{#1}\@@href}%
\providecommand \@@href[1]{\endgroup#1\@@endlink}%
\providecommand \@sanitize@url [0]{\catcode `\\12\catcode `\$12\catcode `\&12\catcode `\#12\catcode `\^12\catcode `\_12\catcode `\%12\relax}%
\providecommand \@@startlink[1]{}%
\providecommand \@@endlink[0]{}%
\providecommand \url  [0]{\begingroup\@sanitize@url \@url }%
\providecommand \@url [1]{\endgroup\@href {#1}{\urlprefix }}%
\providecommand \urlprefix  [0]{URL }%
\providecommand \Eprint [0]{\href }%
\providecommand \doibase [0]{https://doi.org/}%
\providecommand \selectlanguage [0]{\@gobble}%
\providecommand \bibinfo  [0]{\@secondoftwo}%
\providecommand \bibfield  [0]{\@secondoftwo}%
\providecommand \translation [1]{[#1]}%
\providecommand \BibitemOpen [0]{}%
\providecommand \bibitemStop [0]{}%
\providecommand \bibitemNoStop [0]{.\EOS\space}%
\providecommand \EOS [0]{\spacefactor3000\relax}%
\providecommand \BibitemShut  [1]{\csname bibitem#1\endcsname}%
\let\auto@bib@innerbib\@empty
\bibitem [{\citenamefont {Gambetta}(2020)}]{gambetta2020ibm}%
  \BibitemOpen
  \bibfield  {author} {\bibinfo {author} {\bibfnamefont {J.}~\bibnamefont {Gambetta}},\ }\bibfield  {title} {\bibinfo {title} {Ibm’s roadmap for scaling quantum technology},\ }\href@noop {} {\bibfield  {journal} {\bibinfo  {journal} {IBM Research Blog (September 2020)}\ } (\bibinfo {year} {2020})}\BibitemShut {NoStop}%
\bibitem [{\citenamefont {Krinner}\ \emph {et~al.}(2019)\citenamefont {Krinner}, \citenamefont {Storz}, \citenamefont {Kurpiers}, \citenamefont {Magnard}, \citenamefont {Heinsoo}, \citenamefont {Keller}, \citenamefont {Lütolf}, \citenamefont {Eichler},\ and\ \citenamefont {Wallraff}}]{krinner_engineering_2019}%
  \BibitemOpen
  \bibfield  {author} {\bibinfo {author} {\bibfnamefont {S.}~\bibnamefont {Krinner}}, \bibinfo {author} {\bibfnamefont {S.}~\bibnamefont {Storz}}, \bibinfo {author} {\bibfnamefont {P.}~\bibnamefont {Kurpiers}}, \bibinfo {author} {\bibfnamefont {P.}~\bibnamefont {Magnard}}, \bibinfo {author} {\bibfnamefont {J.}~\bibnamefont {Heinsoo}}, \bibinfo {author} {\bibfnamefont {R.}~\bibnamefont {Keller}}, \bibinfo {author} {\bibfnamefont {J.}~\bibnamefont {Lütolf}}, \bibinfo {author} {\bibfnamefont {C.}~\bibnamefont {Eichler}},\ and\ \bibinfo {author} {\bibfnamefont {A.}~\bibnamefont {Wallraff}},\ }\bibfield  {title} {\bibinfo {title} {Engineering cryogenic setups for 100-qubit scale superconducting circuit systems},\ }\href {https://doi.org/10.1140/epjqt/s40507-019-0072-0} {\bibfield  {journal} {\bibinfo  {journal} {EPJ Quantum Technology}\ }\textbf {\bibinfo {volume} {6}},\ \bibinfo {pages} {2} (\bibinfo {year} {2019})}\BibitemShut {NoStop}%
\bibitem [{\citenamefont {Savin}\ \emph {et~al.}(2006)\citenamefont {Savin}, \citenamefont {Pekola}, \citenamefont {Holmqvist}, \citenamefont {Hassel}, \citenamefont {Grönberg}, \citenamefont {Helistö},\ and\ \citenamefont {Kidiyarova-Shevchenko}}]{savin_high-resolution_2006}%
  \BibitemOpen
  \bibfield  {author} {\bibinfo {author} {\bibfnamefont {A.~M.}\ \bibnamefont {Savin}}, \bibinfo {author} {\bibfnamefont {J.~P.}\ \bibnamefont {Pekola}}, \bibinfo {author} {\bibfnamefont {T.}~\bibnamefont {Holmqvist}}, \bibinfo {author} {\bibfnamefont {J.}~\bibnamefont {Hassel}}, \bibinfo {author} {\bibfnamefont {L.}~\bibnamefont {Grönberg}}, \bibinfo {author} {\bibfnamefont {P.}~\bibnamefont {Helistö}},\ and\ \bibinfo {author} {\bibfnamefont {A.}~\bibnamefont {Kidiyarova-Shevchenko}},\ }\bibfield  {title} {\bibinfo {title} {High-resolution superconducting single-flux quantum comparator for sub-{Kelvin} temperatures},\ }\href {https://doi.org/10.1063/1.2357858} {\bibfield  {journal} {\bibinfo  {journal} {Applied Physics Letters}\ }\textbf {\bibinfo {volume} {89}},\ \bibinfo {pages} {133505} (\bibinfo {year} {2006})}\BibitemShut {NoStop}%
\bibitem [{\citenamefont {Intiso}\ \emph {et~al.}(2006)\citenamefont {Intiso}, \citenamefont {Pekola}, \citenamefont {Savin}, \citenamefont {Devyatov},\ and\ \citenamefont {Kidiyarova-Shevchenko}}]{intiso_rapid_2006}%
  \BibitemOpen
  \bibfield  {author} {\bibinfo {author} {\bibfnamefont {S.}~\bibnamefont {Intiso}}, \bibinfo {author} {\bibfnamefont {J.}~\bibnamefont {Pekola}}, \bibinfo {author} {\bibfnamefont {A.}~\bibnamefont {Savin}}, \bibinfo {author} {\bibfnamefont {Y.}~\bibnamefont {Devyatov}},\ and\ \bibinfo {author} {\bibfnamefont {A.}~\bibnamefont {Kidiyarova-Shevchenko}},\ }\bibfield  {title} {\bibinfo {title} {Rapid single-flux-quantum circuits for low noise {mK} operation},\ }\href {https://doi.org/10.1088/0953-2048/19/5/S36} {\bibfield  {journal} {\bibinfo  {journal} {Superconductor Science and Technology}\ }\textbf {\bibinfo {volume} {19}},\ \bibinfo {pages} {S335} (\bibinfo {year} {2006})}\BibitemShut {NoStop}%
\bibitem [{\citenamefont {Tolpygo}\ \emph {et~al.}(2015)\citenamefont {Tolpygo}, \citenamefont {Bolkhovsky}, \citenamefont {Weir}, \citenamefont {Johnson}, \citenamefont {Gouker},\ and\ \citenamefont {Oliver}}]{tolpygo_fabrication_2015}%
  \BibitemOpen
  \bibfield  {author} {\bibinfo {author} {\bibfnamefont {S.~K.}\ \bibnamefont {Tolpygo}}, \bibinfo {author} {\bibfnamefont {V.}~\bibnamefont {Bolkhovsky}}, \bibinfo {author} {\bibfnamefont {T.~J.}\ \bibnamefont {Weir}}, \bibinfo {author} {\bibfnamefont {L.~M.}\ \bibnamefont {Johnson}}, \bibinfo {author} {\bibfnamefont {M.~A.}\ \bibnamefont {Gouker}},\ and\ \bibinfo {author} {\bibfnamefont {W.~D.}\ \bibnamefont {Oliver}},\ }\bibfield  {title} {\bibinfo {title} {Fabrication process and properties of fully-planarized deep-submicron nb/al– $\hbox{AlO}_{\rm x}\hbox{/Nb} $ josephson junctions for vlsi circuits},\ }\href {https://doi.org/10.1109/TASC.2014.2374836} {\bibfield  {journal} {\bibinfo  {journal} {IEEE Transactions on Applied Superconductivity}\ }\textbf {\bibinfo {volume} {25}},\ \bibinfo {pages} {1} (\bibinfo {year} {2015})}\BibitemShut {NoStop}%
\bibitem [{\citenamefont {Wustmann}\ and\ \citenamefont {Osborn}(2020)}]{wustmann_reversible_2020}%
  \BibitemOpen
  \bibfield  {author} {\bibinfo {author} {\bibfnamefont {W.}~\bibnamefont {Wustmann}}\ and\ \bibinfo {author} {\bibfnamefont {K.~D.}\ \bibnamefont {Osborn}},\ }\bibfield  {title} {\bibinfo {title} {Reversible {Fluxon} {Logic}: {Topological} particles allow ballistic gates along {1D} paths},\ }\href {https://doi.org/10.1103/PhysRevB.101.014516} {\bibfield  {journal} {\bibinfo  {journal} {Physical Review B}\ }\textbf {\bibinfo {volume} {101}},\ \bibinfo {pages} {014516} (\bibinfo {year} {2020})},\ \bibinfo {note} {arXiv: 1711.04339}\BibitemShut {NoStop}%
\bibitem [{\citenamefont {Yan}\ \emph {et~al.}(2021)\citenamefont {Yan}, \citenamefont {Hassel}, \citenamefont {Vesterinen}, \citenamefont {Zhang}, \citenamefont {Ikonen}, \citenamefont {Grönberg}, \citenamefont {Goetz},\ and\ \citenamefont {Möttönen}}]{yan_low-noise_2021}%
  \BibitemOpen
  \bibfield  {author} {\bibinfo {author} {\bibfnamefont {C.}~\bibnamefont {Yan}}, \bibinfo {author} {\bibfnamefont {J.}~\bibnamefont {Hassel}}, \bibinfo {author} {\bibfnamefont {V.}~\bibnamefont {Vesterinen}}, \bibinfo {author} {\bibfnamefont {J.}~\bibnamefont {Zhang}}, \bibinfo {author} {\bibfnamefont {J.}~\bibnamefont {Ikonen}}, \bibinfo {author} {\bibfnamefont {L.}~\bibnamefont {Grönberg}}, \bibinfo {author} {\bibfnamefont {J.}~\bibnamefont {Goetz}},\ and\ \bibinfo {author} {\bibfnamefont {M.}~\bibnamefont {Möttönen}},\ }\bibfield  {title} {\bibinfo {title} {A low-noise on-chip coherent microwave source},\ }\href {https://doi.org/10.1038/s41928-021-00680-z} {\bibfield  {journal} {\bibinfo  {journal} {Nature Electronics}\ }\textbf {\bibinfo {volume} {4}},\ \bibinfo {pages} {885} (\bibinfo {year} {2021})}\BibitemShut {NoStop}%
\bibitem [{\citenamefont {Kirichenko}\ \emph {et~al.}(2011)\citenamefont {Kirichenko}, \citenamefont {Sarwana},\ and\ \citenamefont {Kirichenko}}]{kirichenko_zero_2011}%
  \BibitemOpen
  \bibfield  {author} {\bibinfo {author} {\bibfnamefont {D.~E.}\ \bibnamefont {Kirichenko}}, \bibinfo {author} {\bibfnamefont {S.}~\bibnamefont {Sarwana}},\ and\ \bibinfo {author} {\bibfnamefont {A.~F.}\ \bibnamefont {Kirichenko}},\ }\bibfield  {title} {\bibinfo {title} {Zero {Static} {Power} {Dissipation} {Biasing} of {RSFQ} {Circuits}},\ }\href {https://doi.org/10.1109/TASC.2010.2098432} {\bibfield  {journal} {\bibinfo  {journal} {IEEE Transactions on Applied Superconductivity}\ }\textbf {\bibinfo {volume} {21}},\ \bibinfo {pages} {776} (\bibinfo {year} {2011})}\BibitemShut {NoStop}%
\bibitem [{\citenamefont {{K.Likharev and V.K. Semenov}}(1991)}]{Likharev_RSFQ_logic}%
  \BibitemOpen
  \bibfield  {author} {\bibinfo {author} {\bibnamefont {{K.Likharev and V.K. Semenov}}},\ }\bibfield  {title} {\bibinfo {title} {Rsfq logic/memory family: a new josephson-junction technology for sub-terahertz-clock-frequency digital systems},\ }\href {https://doi.org/10.1109/77.80745} {\bibfield  {journal} {\bibinfo  {journal} {IEEE Transactions on Applied Superconductivity}\ }\textbf {\bibinfo {volume} {1}},\ \bibinfo {pages} {3} (\bibinfo {year} {1991})}\BibitemShut {NoStop}%
\bibitem [{\citenamefont {Fedorov}\ \emph {et~al.}(2014)\citenamefont {Fedorov}, \citenamefont {Shcherbakova}, \citenamefont {Wolf}, \citenamefont {Beckmann},\ and\ \citenamefont {Ustinov}}]{fedorov_fluxon_2014}%
  \BibitemOpen
  \bibfield  {author} {\bibinfo {author} {\bibfnamefont {K.~G.}\ \bibnamefont {Fedorov}}, \bibinfo {author} {\bibfnamefont {A.~V.}\ \bibnamefont {Shcherbakova}}, \bibinfo {author} {\bibfnamefont {M.~J.}\ \bibnamefont {Wolf}}, \bibinfo {author} {\bibfnamefont {D.}~\bibnamefont {Beckmann}},\ and\ \bibinfo {author} {\bibfnamefont {A.~V.}\ \bibnamefont {Ustinov}},\ }\bibfield  {title} {\bibinfo {title} {{Fluxon {Readout} of a {Superconducting} {Qubit}}},\ }\href {https://doi.org/10.1103/PhysRevLett.112.160502} {\bibfield  {journal} {\bibinfo  {journal} {Physical Review Letters}\ }\textbf {\bibinfo {volume} {112}},\ \bibinfo {pages} {160502} (\bibinfo {year} {2014})}\BibitemShut {NoStop}%
\bibitem [{\citenamefont {Leonard}\ \emph {et~al.}(2019)\citenamefont {Leonard}, \citenamefont {Beck}, \citenamefont {Nelson}, \citenamefont {Christensen}, \citenamefont {Thorbeck}, \citenamefont {Howington}, \citenamefont {Opremcak}, \citenamefont {Pechenezhskiy}, \citenamefont {Dodge}, \citenamefont {Dupuis}, \citenamefont {Hutchings}, \citenamefont {Ku}, \citenamefont {Schlenker}, \citenamefont {Suttle}, \citenamefont {Wilen}, \citenamefont {Zhu}, \citenamefont {Vavilov}, \citenamefont {Plourde},\ and\ \citenamefont {McDermott}}]{Leonard_SFQ_Control}%
  \BibitemOpen
  \bibfield  {author} {\bibinfo {author} {\bibfnamefont {E.}~\bibnamefont {Leonard}}, \bibinfo {author} {\bibfnamefont {M.~A.}\ \bibnamefont {Beck}}, \bibinfo {author} {\bibfnamefont {J.}~\bibnamefont {Nelson}}, \bibinfo {author} {\bibfnamefont {B.}~\bibnamefont {Christensen}}, \bibinfo {author} {\bibfnamefont {T.}~\bibnamefont {Thorbeck}}, \bibinfo {author} {\bibfnamefont {C.}~\bibnamefont {Howington}}, \bibinfo {author} {\bibfnamefont {A.}~\bibnamefont {Opremcak}}, \bibinfo {author} {\bibfnamefont {I.}~\bibnamefont {Pechenezhskiy}}, \bibinfo {author} {\bibfnamefont {K.}~\bibnamefont {Dodge}}, \bibinfo {author} {\bibfnamefont {N.}~\bibnamefont {Dupuis}}, \bibinfo {author} {\bibfnamefont {M.}~\bibnamefont {Hutchings}}, \bibinfo {author} {\bibfnamefont {J.}~\bibnamefont {Ku}}, \bibinfo {author} {\bibfnamefont {F.}~\bibnamefont {Schlenker}}, \bibinfo {author} {\bibfnamefont {J.}~\bibnamefont {Suttle}}, \bibinfo {author} {\bibfnamefont {C.}~\bibnamefont {Wilen}}, \bibinfo {author} {\bibfnamefont
  {S.}~\bibnamefont {Zhu}}, \bibinfo {author} {\bibfnamefont {M.}~\bibnamefont {Vavilov}}, \bibinfo {author} {\bibfnamefont {B.}~\bibnamefont {Plourde}},\ and\ \bibinfo {author} {\bibfnamefont {R.}~\bibnamefont {McDermott}},\ }\bibfield  {title} {\bibinfo {title} {Digital coherent control of a superconducting qubit},\ }\href {https://doi.org/10.1103/PhysRevApplied.11.014009} {\bibfield  {journal} {\bibinfo  {journal} {Phys. Rev. Appl.}\ }\textbf {\bibinfo {volume} {11}},\ \bibinfo {pages} {014009} (\bibinfo {year} {2019})}\BibitemShut {NoStop}%
\bibitem [{\citenamefont {McDermott}\ \emph {et~al.}(2018)\citenamefont {McDermott}, \citenamefont {Vavilov}, \citenamefont {Plourde}, \citenamefont {Wilhelm}, \citenamefont {Liebermann}, \citenamefont {Mukhanov},\ and\ \citenamefont {Ohki}}]{mcdermott_quantumclassical_2018}%
  \BibitemOpen
  \bibfield  {author} {\bibinfo {author} {\bibfnamefont {R.}~\bibnamefont {McDermott}}, \bibinfo {author} {\bibfnamefont {M.~G.}\ \bibnamefont {Vavilov}}, \bibinfo {author} {\bibfnamefont {B.~L.~T.}\ \bibnamefont {Plourde}}, \bibinfo {author} {\bibfnamefont {F.~K.}\ \bibnamefont {Wilhelm}}, \bibinfo {author} {\bibfnamefont {P.~J.}\ \bibnamefont {Liebermann}}, \bibinfo {author} {\bibfnamefont {O.~A.}\ \bibnamefont {Mukhanov}},\ and\ \bibinfo {author} {\bibfnamefont {T.~A.}\ \bibnamefont {Ohki}},\ }\bibfield  {title} {\bibinfo {title} {Quantum–classical interface based on single flux quantum digital logic},\ }\href {https://doi.org/10.1088/2058-9565/aaa3a0} {\bibfield  {journal} {\bibinfo  {journal} {Quantum Science and Technology}\ }\textbf {\bibinfo {volume} {3}},\ \bibinfo {pages} {024004} (\bibinfo {year} {2018})}\BibitemShut {NoStop}%
\bibitem [{\citenamefont {McLaughlin}\ and\ \citenamefont {Scott}(1978)}]{mclaughlin_perturbation_1978}%
  \BibitemOpen
  \bibfield  {author} {\bibinfo {author} {\bibfnamefont {D.~W.}\ \bibnamefont {McLaughlin}}\ and\ \bibinfo {author} {\bibfnamefont {A.~C.}\ \bibnamefont {Scott}},\ }\bibfield  {title} {\bibinfo {title} {Perturbation analysis of fluxon dynamics},\ }\href {https://doi.org/10.1103/PhysRevA.18.1652} {\bibfield  {journal} {\bibinfo  {journal} {Physical Review A}\ }\textbf {\bibinfo {volume} {18}},\ \bibinfo {pages} {1652} (\bibinfo {year} {1978})}\BibitemShut {NoStop}%
\bibitem [{\citenamefont {Olsen}\ and\ \citenamefont {Samuelsen}(1981)}]{olsen_reflection_1981}%
  \BibitemOpen
  \bibfield  {author} {\bibinfo {author} {\bibfnamefont {O.~H.}\ \bibnamefont {Olsen}}\ and\ \bibinfo {author} {\bibfnamefont {M.~R.}\ \bibnamefont {Samuelsen}},\ }\bibfield  {title} {\bibinfo {title} {Reflection of sine‐{Gordon} breathers},\ }\href {https://doi.org/10.1063/1.329026} {\bibfield  {journal} {\bibinfo  {journal} {Journal of Applied Physics}\ }\textbf {\bibinfo {volume} {52}},\ \bibinfo {pages} {2913} (\bibinfo {year} {1981})}\BibitemShut {NoStop}%
\bibitem [{\citenamefont {Costabile}\ \emph {et~al.}(1978)\citenamefont {Costabile}, \citenamefont {Parmentier}, \citenamefont {Savo}, \citenamefont {McLaughlin},\ and\ \citenamefont {Scott}}]{costabile_exact_1978}%
  \BibitemOpen
  \bibfield  {author} {\bibinfo {author} {\bibfnamefont {G.}~\bibnamefont {Costabile}}, \bibinfo {author} {\bibfnamefont {R.~D.}\ \bibnamefont {Parmentier}}, \bibinfo {author} {\bibfnamefont {B.}~\bibnamefont {Savo}}, \bibinfo {author} {\bibfnamefont {D.~W.}\ \bibnamefont {McLaughlin}},\ and\ \bibinfo {author} {\bibfnamefont {A.~C.}\ \bibnamefont {Scott}},\ }\bibfield  {title} {\bibinfo {title} {Exact solutions of the sine‐{Gordon} equation describing oscillations in a long (but finite) {Josephson} junction},\ }\href {https://doi.org/10.1063/1.90113} {\bibfield  {journal} {\bibinfo  {journal} {Applied Physics Letters}\ }\textbf {\bibinfo {volume} {32}},\ \bibinfo {pages} {587} (\bibinfo {year} {1978})}\BibitemShut {NoStop}%
\bibitem [{\citenamefont {Christiansen}\ and\ \citenamefont {Olsen}(1980)}]{christiansen_reflection_1980}%
  \BibitemOpen
  \bibfield  {author} {\bibinfo {author} {\bibfnamefont {P.}~\bibnamefont {Christiansen}}\ and\ \bibinfo {author} {\bibfnamefont {O.}~\bibnamefont {Olsen}},\ }\bibfield  {title} {\bibinfo {title} {Reflection of fluxons on a {Josephson} line cavity},\ }\href {https://doi.org/10.1016/0167-2789(80)90021-4} {\bibfield  {journal} {\bibinfo  {journal} {Physica D: Nonlinear Phenomena}\ }\textbf {\bibinfo {volume} {1}},\ \bibinfo {pages} {412} (\bibinfo {year} {1980})}\BibitemShut {NoStop}%
\bibitem [{\citenamefont {Scott}\ \emph {et~al.}(1973)\citenamefont {Scott}, \citenamefont {Chu},\ and\ \citenamefont {McLaughlin}}]{Scott_solitons_1973}%
  \BibitemOpen
  \bibfield  {author} {\bibinfo {author} {\bibfnamefont {A.}~\bibnamefont {Scott}}, \bibinfo {author} {\bibfnamefont {F.}~\bibnamefont {Chu}},\ and\ \bibinfo {author} {\bibfnamefont {D.}~\bibnamefont {McLaughlin}},\ }\bibfield  {title} {\bibinfo {title} {The soliton: A new concept in applied science},\ }\href {https://doi.org/10.1109/PROC.1973.9296} {\bibfield  {journal} {\bibinfo  {journal} {Proceedings of the IEEE}\ }\textbf {\bibinfo {volume} {61}},\ \bibinfo {pages} {1443} (\bibinfo {year} {1973})}\BibitemShut {NoStop}%
\bibitem [{\citenamefont {Zakharov}\ and\ \citenamefont {Shabat}(1971)}]{ZS71}%
  \BibitemOpen
  \bibfield  {author} {\bibinfo {author} {\bibfnamefont {V.~E.}\ \bibnamefont {Zakharov}}\ and\ \bibinfo {author} {\bibfnamefont {A.~B.}\ \bibnamefont {Shabat}},\ }\bibfield  {title} {\bibinfo {title} {Exact theory of two-dimensional self-focusing and one-dimensional self-modulation of waves in nonlinear media},\ }\href@noop {} {\bibfield  {journal} {\bibinfo  {journal} {Zh. Eksp. Teor. Fiz.}\ }\textbf {\bibinfo {volume} {61}},\ \bibinfo {pages} {118} (\bibinfo {year} {1971})},\ \translation{Sov. Phys. JETP \textbf{34}, 62 (1972)}\BibitemShut {NoStop}%
\bibitem [{\citenamefont {Likharev}(1986)}]{Likharev_JJ_dynamics}%
  \BibitemOpen
  \bibfield  {author} {\bibinfo {author} {\bibfnamefont {K.}~\bibnamefont {Likharev}},\ }\href@noop {} {\emph {\bibinfo {title} {Dynamics of Josephson Junctions and Circuits}}}\ (\bibinfo  {publisher} {Routledge},\ \bibinfo {address} {London},\ \bibinfo {year} {1986})\BibitemShut {NoStop}%
\bibitem [{\citenamefont {Remoissenet}(1996)}]{remoissenet_waves_1996}%
  \BibitemOpen
  \bibfield  {author} {\bibinfo {author} {\bibfnamefont {M.}~\bibnamefont {Remoissenet}},\ }\href {http://public.eblib.com/choice/PublicFullRecord.aspx?p=6555146} {\emph {\bibinfo {title} {Waves {Called} {Solitons} {Concepts} and {Experiments}.}}}\ (\bibinfo  {publisher} {Springer Berlin / Heidelberg},\ \bibinfo {address} {Berlin, Heidelberg},\ \bibinfo {year} {1996})\ \bibinfo {note} {oCLC: 1255224501}\BibitemShut {NoStop}%
\bibitem [{\citenamefont {McCumber}(1968)}]{McCumber_OG}%
  \BibitemOpen
  \bibfield  {author} {\bibinfo {author} {\bibfnamefont {D.~E.}\ \bibnamefont {McCumber}},\ }\bibfield  {title} {\bibinfo {title} {{Effect of ac Impedance on dc Voltage‐Current Characteristics of Superconductor Weak‐Link Junctions}},\ }\href {https://doi.org/10.1063/1.1656743} {\bibfield  {journal} {\bibinfo  {journal} {Journal of Applied Physics}\ }\textbf {\bibinfo {volume} {39}},\ \bibinfo {pages} {3113} (\bibinfo {year} {1968})}\BibitemShut {NoStop}%
\bibitem [{\citenamefont {Stewart}(1968)}]{Stewart_OG}%
  \BibitemOpen
  \bibfield  {author} {\bibinfo {author} {\bibfnamefont {W.~C.}\ \bibnamefont {Stewart}},\ }\bibfield  {title} {\bibinfo {title} {{CURRENT‐VOLTAGE CHARACTERISTICS OF JOSEPHSON JUNCTIONS}},\ }\href {https://doi.org/10.1063/1.1651991} {\bibfield  {journal} {\bibinfo  {journal} {Applied Physics Letters}\ }\textbf {\bibinfo {volume} {12}},\ \bibinfo {pages} {277} (\bibinfo {year} {1968})}\BibitemShut {NoStop}%
\bibitem [{\citenamefont {Kivshar}\ and\ \citenamefont {Malomed}(1989)}]{kivshar_dynamics_1989}%
  \BibitemOpen
  \bibfield  {author} {\bibinfo {author} {\bibfnamefont {Y.~S.}\ \bibnamefont {Kivshar}}\ and\ \bibinfo {author} {\bibfnamefont {B.~A.}\ \bibnamefont {Malomed}},\ }\bibfield  {title} {\bibinfo {title} {Dynamics of solitons in nearly integrable systems},\ }\href {https://doi.org/10.1103/RevModPhys.61.763} {\bibfield  {journal} {\bibinfo  {journal} {Reviews of Modern Physics}\ }\textbf {\bibinfo {volume} {61}},\ \bibinfo {pages} {763} (\bibinfo {year} {1989})}\BibitemShut {NoStop}%
\bibitem [{\citenamefont {Peyrard}\ and\ \citenamefont {Kruskal}(1984)}]{peyrard_kink_1984}%
  \BibitemOpen
  \bibfield  {author} {\bibinfo {author} {\bibfnamefont {M.}~\bibnamefont {Peyrard}}\ and\ \bibinfo {author} {\bibfnamefont {M.~D.}\ \bibnamefont {Kruskal}},\ }\bibfield  {title} {\bibinfo {title} {Kink dynamics in the highly discrete sine-{Gordon} system},\ }\href {https://doi.org/10.1016/0167-2789(84)90006-X} {\bibfield  {journal} {\bibinfo  {journal} {Physica D: Nonlinear Phenomena}\ }\textbf {\bibinfo {volume} {14}},\ \bibinfo {pages} {88} (\bibinfo {year} {1984})}\BibitemShut {NoStop}%
\bibitem [{\citenamefont {Kurin}\ and\ \citenamefont {Yulin}(1997)}]{kurin_radiation_1997}%
  \BibitemOpen
  \bibfield  {author} {\bibinfo {author} {\bibfnamefont {V.~V.}\ \bibnamefont {Kurin}}\ and\ \bibinfo {author} {\bibfnamefont {A.~V.}\ \bibnamefont {Yulin}},\ }\bibfield  {title} {\bibinfo {title} {Radiation of linear waves by solitons in a {Josephson} transmission line with dispersion},\ }\href {https://doi.org/10.1103/PhysRevB.55.11659} {\bibfield  {journal} {\bibinfo  {journal} {Physical Review B}\ }\textbf {\bibinfo {volume} {55}},\ \bibinfo {pages} {11659} (\bibinfo {year} {1997})}\BibitemShut {NoStop}%
\bibitem [{\citenamefont {Osborn}\ and\ \citenamefont {Wustmann}(2023)}]{osborn_asynchronous_2022}%
  \BibitemOpen
  \bibfield  {author} {\bibinfo {author} {\bibfnamefont {K.}~\bibnamefont {Osborn}}\ and\ \bibinfo {author} {\bibfnamefont {W.}~\bibnamefont {Wustmann}},\ }\bibfield  {title} {\bibinfo {title} {Asynchronous reversible computing unveiled using ballistic shift registers},\ }\href {https://doi.org/10.1103/PhysRevApplied.19.054034} {\bibfield  {journal} {\bibinfo  {journal} {Phys. Rev. Appl.}\ }\textbf {\bibinfo {volume} {19}},\ \bibinfo {pages} {054034} (\bibinfo {year} {2023})}\BibitemShut {NoStop}%
\bibitem [{\citenamefont {Darula}\ and\ \citenamefont {Kedro}(1990)}]{darula_dynamic_1990}%
  \BibitemOpen
  \bibfield  {author} {\bibinfo {author} {\bibfnamefont {M.}~\bibnamefont {Darula}}\ and\ \bibinfo {author} {\bibfnamefont {M.}~\bibnamefont {Kedro}},\ }\bibfield  {title} {\bibinfo {title} {Dynamic reduction of the critical current in a {Josephson} junction},\ }\href {https://doi.org/10.1007/BF00683311} {\bibfield  {journal} {\bibinfo  {journal} {Journal of Low Temperature Physics}\ }\textbf {\bibinfo {volume} {78}},\ \bibinfo {pages} {287} (\bibinfo {year} {1990})}\BibitemShut {NoStop}%
\bibitem [{\citenamefont {Currie}\ \emph {et~al.}(1977)\citenamefont {Currie}, \citenamefont {Trullinger}, \citenamefont {Bishop},\ and\ \citenamefont {Krumhansl}}]{currie_numerical_1977}%
  \BibitemOpen
  \bibfield  {author} {\bibinfo {author} {\bibfnamefont {J.~F.}\ \bibnamefont {Currie}}, \bibinfo {author} {\bibfnamefont {S.~E.}\ \bibnamefont {Trullinger}}, \bibinfo {author} {\bibfnamefont {A.~R.}\ \bibnamefont {Bishop}},\ and\ \bibinfo {author} {\bibfnamefont {J.~A.}\ \bibnamefont {Krumhansl}},\ }\bibfield  {title} {\bibinfo {title} {Numerical simulation of sine-{Gordon} soliton dynamics in the presence of perturbations},\ }\href {https://doi.org/10.1103/PhysRevB.15.5567} {\bibfield  {journal} {\bibinfo  {journal} {Physical Review B}\ }\textbf {\bibinfo {volume} {15}},\ \bibinfo {pages} {5567} (\bibinfo {year} {1977})}\BibitemShut {NoStop}%
\bibitem [{\citenamefont {Polonsky}\ \emph {et~al.}(1993)\citenamefont {Polonsky}, \citenamefont {Semenov},\ and\ \citenamefont {Schneider}}]{polonsky_transmission_nodate}%
  \BibitemOpen
  \bibfield  {author} {\bibinfo {author} {\bibfnamefont {S.}~\bibnamefont {Polonsky}}, \bibinfo {author} {\bibfnamefont {V.}~\bibnamefont {Semenov}},\ and\ \bibinfo {author} {\bibfnamefont {D.}~\bibnamefont {Schneider}},\ }\bibfield  {title} {\bibinfo {title} {Transmission of single-flux-quantum pulses along superconducting microstrip lines},\ }\href {https://doi.org/10.1109/77.233525} {\bibfield  {journal} {\bibinfo  {journal} {IEEE Transactions on Applied Superconductivity}\ }\textbf {\bibinfo {volume} {3}},\ \bibinfo {pages} {2598} (\bibinfo {year} {1993})}\BibitemShut {NoStop}%
\bibitem [{\citenamefont {Wildermuth}\ \emph {et~al.}(2022)\citenamefont {Wildermuth}, \citenamefont {Powalla}, \citenamefont {Voss}, \citenamefont {Schön}, \citenamefont {Schneider}, \citenamefont {Fistul}, \citenamefont {Rotzinger},\ and\ \citenamefont {Ustinov}}]{fluxoids_ustinov}%
  \BibitemOpen
  \bibfield  {author} {\bibinfo {author} {\bibfnamefont {M.}~\bibnamefont {Wildermuth}}, \bibinfo {author} {\bibfnamefont {L.}~\bibnamefont {Powalla}}, \bibinfo {author} {\bibfnamefont {J.~N.}\ \bibnamefont {Voss}}, \bibinfo {author} {\bibfnamefont {Y.}~\bibnamefont {Schön}}, \bibinfo {author} {\bibfnamefont {A.}~\bibnamefont {Schneider}}, \bibinfo {author} {\bibfnamefont {M.~V.}\ \bibnamefont {Fistul}}, \bibinfo {author} {\bibfnamefont {H.}~\bibnamefont {Rotzinger}},\ and\ \bibinfo {author} {\bibfnamefont {A.~V.}\ \bibnamefont {Ustinov}},\ }\bibfield  {title} {\bibinfo {title} {{Fluxons in high-impedance long Josephson junctions}},\ }\href {https://doi.org/10.1063/5.0082197} {\bibfield  {journal} {\bibinfo  {journal} {Applied Physics Letters}\ }\textbf {\bibinfo {volume} {120}},\ \bibinfo {pages} {112601} (\bibinfo {year} {2022})}\BibitemShut {NoStop}%
\bibitem [{\citenamefont {Krantz}\ \emph {et~al.}(2019)\citenamefont {Krantz}, \citenamefont {Kjaergaard}, \citenamefont {Yan}, \citenamefont {Orlando}, \citenamefont {Gustavsson},\ and\ \citenamefont {Oliver}}]{krantz_quantum_2019}%
  \BibitemOpen
  \bibfield  {author} {\bibinfo {author} {\bibfnamefont {P.}~\bibnamefont {Krantz}}, \bibinfo {author} {\bibfnamefont {M.}~\bibnamefont {Kjaergaard}}, \bibinfo {author} {\bibfnamefont {F.}~\bibnamefont {Yan}}, \bibinfo {author} {\bibfnamefont {T.~P.}\ \bibnamefont {Orlando}}, \bibinfo {author} {\bibfnamefont {S.}~\bibnamefont {Gustavsson}},\ and\ \bibinfo {author} {\bibfnamefont {W.~D.}\ \bibnamefont {Oliver}},\ }\bibfield  {title} {\bibinfo {title} {A {Quantum} {Engineer}'s {Guide} to {Superconducting} {Qubits}},\ }\href {https://doi.org/10.1063/1.5089550} {\bibfield  {journal} {\bibinfo  {journal} {Applied Physics Reviews}\ }\textbf {\bibinfo {volume} {6}},\ \bibinfo {pages} {021318} (\bibinfo {year} {2019})},\ \bibinfo {note} {arXiv: 1904.06560}\BibitemShut {NoStop}%
\bibitem [{\citenamefont {Koch}\ \emph {et~al.}(2007)\citenamefont {Koch}, \citenamefont {Yu}, \citenamefont {Gambetta}, \citenamefont {Houck}, \citenamefont {Schuster}, \citenamefont {Majer}, \citenamefont {Blais}, \citenamefont {Devoret}, \citenamefont {Girvin},\ and\ \citenamefont {Schoelkopf}}]{Transmon_OG_Koch}%
  \BibitemOpen
  \bibfield  {author} {\bibinfo {author} {\bibfnamefont {J.}~\bibnamefont {Koch}}, \bibinfo {author} {\bibfnamefont {T.~M.}\ \bibnamefont {Yu}}, \bibinfo {author} {\bibfnamefont {J.}~\bibnamefont {Gambetta}}, \bibinfo {author} {\bibfnamefont {A.~A.}\ \bibnamefont {Houck}}, \bibinfo {author} {\bibfnamefont {D.~I.}\ \bibnamefont {Schuster}}, \bibinfo {author} {\bibfnamefont {J.}~\bibnamefont {Majer}}, \bibinfo {author} {\bibfnamefont {A.}~\bibnamefont {Blais}}, \bibinfo {author} {\bibfnamefont {M.~H.}\ \bibnamefont {Devoret}}, \bibinfo {author} {\bibfnamefont {S.~M.}\ \bibnamefont {Girvin}},\ and\ \bibinfo {author} {\bibfnamefont {R.~J.}\ \bibnamefont {Schoelkopf}},\ }\bibfield  {title} {\bibinfo {title} {Charge-insensitive qubit design derived from the cooper pair box},\ }\href {https://doi.org/10.1103/PhysRevA.76.042319} {\bibfield  {journal} {\bibinfo  {journal} {Phys. Rev. A}\ }\textbf {\bibinfo {volume} {76}},\ \bibinfo {pages} {042319} (\bibinfo {year} {2007})}\BibitemShut {NoStop}%
\bibitem [{\citenamefont {Grnbech-Jensen}\ \emph {et~al.}(1991)\citenamefont {Grnbech-Jensen}, \citenamefont {Lomdahl},\ and\ \citenamefont {Samuelsen}}]{Gronbech_Jensen_fluxon_pendulum}%
  \BibitemOpen
  \bibfield  {author} {\bibinfo {author} {\bibfnamefont {N.}~\bibnamefont {Grnbech-Jensen}}, \bibinfo {author} {\bibfnamefont {P.~S.}\ \bibnamefont {Lomdahl}},\ and\ \bibinfo {author} {\bibfnamefont {M.~R.}\ \bibnamefont {Samuelsen}},\ }\bibfield  {title} {\bibinfo {title} {Bifurcation and chaos in a dc-driven long annular josephson junction},\ }\href {https://doi.org/10.1103/PhysRevB.43.12799} {\bibfield  {journal} {\bibinfo  {journal} {Phys. Rev. B}\ }\textbf {\bibinfo {volume} {43}},\ \bibinfo {pages} {12799} (\bibinfo {year} {1991})}\BibitemShut {NoStop}%
\bibitem [{\citenamefont {Howe}\ \emph {et~al.}(2022)\citenamefont {Howe}, \citenamefont {Castellanos-Beltran}, \citenamefont {Sirois}, \citenamefont {Olaya}, \citenamefont {Biesecker}, \citenamefont {Dresselhaus}, \citenamefont {Benz},\ and\ \citenamefont {Hopkins}}]{JPG_3K}%
  \BibitemOpen
  \bibfield  {author} {\bibinfo {author} {\bibfnamefont {L.}~\bibnamefont {Howe}}, \bibinfo {author} {\bibfnamefont {M.~A.}\ \bibnamefont {Castellanos-Beltran}}, \bibinfo {author} {\bibfnamefont {A.~J.}\ \bibnamefont {Sirois}}, \bibinfo {author} {\bibfnamefont {D.}~\bibnamefont {Olaya}}, \bibinfo {author} {\bibfnamefont {J.}~\bibnamefont {Biesecker}}, \bibinfo {author} {\bibfnamefont {P.~D.}\ \bibnamefont {Dresselhaus}}, \bibinfo {author} {\bibfnamefont {S.~P.}\ \bibnamefont {Benz}},\ and\ \bibinfo {author} {\bibfnamefont {P.~F.}\ \bibnamefont {Hopkins}},\ }\bibfield  {title} {\bibinfo {title} {Digital control of a superconducting qubit using a josephson pulse generator at 3 k},\ }\href {https://doi.org/10.1103/PRXQuantum.3.010350} {\bibfield  {journal} {\bibinfo  {journal} {PRX Quantum}\ }\textbf {\bibinfo {volume} {3}},\ \bibinfo {pages} {010350} (\bibinfo {year} {2022})}\BibitemShut {NoStop}%
\bibitem [{\citenamefont {Gambetta}\ \emph {et~al.}(2006)\citenamefont {Gambetta}, \citenamefont {Blais}, \citenamefont {Schuster}, \citenamefont {Wallraff}, \citenamefont {Frunzio}, \citenamefont {Majer}, \citenamefont {Devoret}, \citenamefont {Girvin},\ and\ \citenamefont {Schoelkopf}}]{Gambetta_qubit_readout_measurement}%
  \BibitemOpen
  \bibfield  {author} {\bibinfo {author} {\bibfnamefont {J.}~\bibnamefont {Gambetta}}, \bibinfo {author} {\bibfnamefont {A.}~\bibnamefont {Blais}}, \bibinfo {author} {\bibfnamefont {D.~I.}\ \bibnamefont {Schuster}}, \bibinfo {author} {\bibfnamefont {A.}~\bibnamefont {Wallraff}}, \bibinfo {author} {\bibfnamefont {L.}~\bibnamefont {Frunzio}}, \bibinfo {author} {\bibfnamefont {J.}~\bibnamefont {Majer}}, \bibinfo {author} {\bibfnamefont {M.~H.}\ \bibnamefont {Devoret}}, \bibinfo {author} {\bibfnamefont {S.~M.}\ \bibnamefont {Girvin}},\ and\ \bibinfo {author} {\bibfnamefont {R.~J.}\ \bibnamefont {Schoelkopf}},\ }\bibfield  {title} {\bibinfo {title} {Qubit-photon interactions in a cavity: Measurement-induced dephasing and number splitting},\ }\href {https://doi.org/10.1103/PhysRevA.74.042318} {\bibfield  {journal} {\bibinfo  {journal} {Phys. Rev. A}\ }\textbf {\bibinfo {volume} {74}},\ \bibinfo {pages} {042318} (\bibinfo {year} {2006})}\BibitemShut {NoStop}%
\bibitem [{\citenamefont {Naaman}\ \emph {et~al.}(2018)\citenamefont {Naaman}, \citenamefont {Strong}, \citenamefont {Ferguson}, \citenamefont {Egan}, \citenamefont {Bailey},\ and\ \citenamefont {Hinkey}}]{Josephson_IQmixer}%
  \BibitemOpen
  \bibfield  {author} {\bibinfo {author} {\bibfnamefont {O.}~\bibnamefont {Naaman}}, \bibinfo {author} {\bibfnamefont {J.}~\bibnamefont {Strong}}, \bibinfo {author} {\bibfnamefont {D.}~\bibnamefont {Ferguson}}, \bibinfo {author} {\bibfnamefont {J.}~\bibnamefont {Egan}}, \bibinfo {author} {\bibfnamefont {N.}~\bibnamefont {Bailey}},\ and\ \bibinfo {author} {\bibfnamefont {R.}~\bibnamefont {Hinkey}},\ }\bibfield  {title} {\bibinfo {title} {Josephson junction microwave modulators},\ }in\ \href {https://doi.org/10.1109/MWSYM.2018.8439493} {\emph {\bibinfo {booktitle} {2018 IEEE/MTT-S International Microwave Symposium - IMS}}}\ (\bibinfo {year} {2018})\ pp.\ \bibinfo {pages} {1003--1005}\BibitemShut {NoStop}%
\end{thebibliography}%

\end{document}